\documentclass{article}
\usepackage{amsfonts}
\usepackage{latexsym}
\newtheorem{theorem}{Theorem}
\newtheorem{lemma}{Lemma}
\newtheorem{definition}{Definition}
\newtheorem{proposition}{Proposition}
\begin{document}
\title{The Stationary Maxwell-Dirac Equations}

\author{Chris Radford\\
School of Mathematical and Computer Sciences\\
         University of New England\\
         Armidale NSW 2351\\
          Australia}

\maketitle

\begin{abstract}The Maxwell-Dirac equations are the equations for electronic 
matter, the ``classical" theory underlying QED. The system combines the Dirac 
equations with the Maxwell equations sourced by the Dirac current.

A stationary Maxwell-Dirac system has $\psi =e^{-iEt}\phi$, 
with $\phi$ independent of $t$. The system is said to be isolated if the 
dependent variables obey quite weak regularity and decay conditions. In this 
paper we prove the following results for isolated, stationary Maxwell-Dirac 
systems,
\begin{itemize}
\item there are no embedded eigenvalues in the essential spectrum, i.e. 
$-m\leq E\leq m$;
\item if $|E|<m$ then the Dirac field decays exponentially as $|x|\to\infty$;
\item if $|E|=m$ then the system is ``asymptotically" static and decays 
exponentially if the total charge is non-zero.
\end{itemize} 
\end{abstract}

\bigskip

\section{Introduction}

The Maxwell-Dirac system consists of the Dirac equation
\begin{equation}\gamma^\alpha (\partial_\alpha -i\,e\,A_\alpha )\psi +
im\psi =0,\label{dir:norm}\end{equation}
with electromagnetic interaction given by the potential $A_\alpha$; and 
the Maxwell equations (sourced by the Dirac current, $j^\alpha$), 
\begin{eqnarray}
&&F_{\alpha\beta}=\partial_\alpha A_\beta -\partial_\beta A_\alpha ,\nonumber\\
&&\partial^\alpha F_{\alpha\beta}= -4\pi ej_\beta =-4\pi e\bar{\psi}
\gamma_\beta 
\psi.\label{max:norm}
\end{eqnarray}

Most studies of the Dirac equation treat the electromagnetic field as given and 
ignore the Dirac current as a source for the Maxwell equations, i.e. these 
treatments ignore the electron ``self-field". A comprehensive survey of these 
results can be found in the book by Thaller\cite{BT:dbook}. This is not 
surprising, inclusion of the electron self-field via the Dirac current leads 
to a very difficult, highly non-linear set of partial differential equations. 
So difficult in fact that the existence theory and solution of the Cauchy 
problem was not completed until 1997
 -- seventy years after Dirac first wrote down his 
equation! In a stunning piece of non-linear analysis
, worked out over an almost twenty year period,  Flato, Simon and Taflin 
(\cite{FST:gexs}) solved the Cauchy problem for small initial data. Other 
contributors to this work on the existence of
 solutions would include Gross \cite{LG:cauch},
Chadam \cite{JC:cauch}, Georgiev \cite{VG:exs}, Esteban {\em et al}
\cite{EGS:stexs}, and Bournaveas \cite{NB:exs}.

There are no known non-trivial, exact solutions to the Maxwell-Dirac equations  
in $1+3$ dimensions -- all known solutions involve some numerical work. These 
solutions do, however, exhibit interesting non-linear behaviour which would 
not have been apparent through perturbation expansions. 
The particular solutions found in \cite{CR:pap1} and \cite{HC:pap2} exhibit just this sort of behaviour -- localisation and
charge screening. See also Das \cite{AD:md} and the more recent work of Finster,
 Smoller and Yau \cite{FSY:mds1}.

Finster, Smoller and Yau also point out in \cite{FSY:mds2} that solving
the system (Einstein-Maxwell-Dirac system in their case) gives, in effect, all
the Feynman diagrams of the quantum field theory, with the exception of the fermionic loop diagrams.
Study of the Maxwell-Dirac system should provide an interesting insight into 
non-perturbative QED.

The aim of the present work is to obtain qualitative information on stationary 
solutions of the Maxwell-Dirac system, in doing so one would hope to be able to 
say something about Maxwell-Dirac models of some simple natural systems 
involving the electron. In fact, it is surprising that so little is understood 
about these systems. There is no Maxwell-Dirac model for a single isolated 
electron. If we compare the situation to that in the other great physical 
theory of the 20th century, General Relativity, the situation could not be more 
stark. There are a host of solutions to Einstein's equations representing 
single, isolated gravitating bodies.

In this article we will prove the following result.

\medskip

\noindent{\bf Main Theorem} {\em A stationary, isolated Maxwell-Dirac system 
has no 
embedded eigenvalues, i.e. $-m\leq E\leq m$. 

\noindent If $|E|<m$ then the Dirac field, 
decays exponentially as $|x|\to\infty$.

\noindent If $|E|=m$ then the system is ``asymptotically static" and 
,if the system has non-vanishing total charge, decays exponentially as 
$|x|\to \infty$.}

\medskip

The content of the theorem will be established through theorems 1 to 5 in 
sections 4 to 7.

The paper is organised as follows. First, we give a brief overview 
of 2-spinor methods applied to the Maxwell-Dirac equations (some details of 
2-spinor calculus may also be found in the appendices). In \S{3} the 
definitions of stationary, isolated systems are given and some simple 
consequences are explored. In \S{4} we address the important 
question of embedded eigenvalues in the spectrum of the Dirac operator, the 
main results being proposition 1 and theorem 1. In \S{5} we examine 
regularity and decay issues with the main results given in theorems 2 and 3. 
The next section, \S{6}, looks at the special case $|E|=m$ and its relation to 
a generalisation of the static systems of \cite{CH:pap3}. In 
\S{7} we prove the exponential decay of the Dirac field in the $|E|=m$ case. 
Finally, in \S{8}, we conclude with a brief discussion of the 
results.

\section{The Maxwell-Dirac Equations}

In this section we give a very brief account of the 2-spinor formulation of the 
Maxwell-Dirac equations, details may be found in \cite{CR:pap1}, some of the 
``mechanics" of the 2-spinor formalism are collected in Appendix A at the end 
 of the 
paper. 

In \cite{CR:pap1} the 2-spinor form of the Dirac equations was employed to solve
(\ref{dir:norm}) for the electromagnetic potential, under the
non-degeneracy condition $j^{\alpha}j_{\alpha}\neq 0$. In terms of 2-spinors 
-- see below -- the non-degeneracy condition can be written as $u_Cv^C\neq 0$, 
since $j^{\alpha}j_{\alpha}=2|u_Cv^C|^2$. 
Requiring $A^{\alpha}$ to be a real 
four-vector gives a set of partial
differential equations in the Dirac field alone, {\em the reality conditions}.

For 2-spinors $u_A$ and $v^B$ (see \cite{PR:book} for an exposition of the 
2-spinor formalism) we have
     \begin{eqnarray}
&&\psi = \left(\begin{array}{c}
     u_A\\
     \overline{v}^{\dot{B}}
     \end{array}
     \right),  \ \ \mbox{with}\nonumber\\
&&u_Cv^C \neq 0 \ \ \mbox{(non-degeneracy)},\nonumber 
      \end{eqnarray}
where $A, B = 0, 1$, $\dot{A},\dot{B} = \dot{0}, \dot{1}$ are two-spinor 
indices.
The Dirac equations are
\begin{eqnarray}
&&(\partial^{A\dot{A}}-
i\,e\,A^{A\dot{A}})u_A+\frac{im}{\sqrt{2}}\overline{v}^{\dot{A}}=0,
\label{dir:2sp}\\
&&(\partial^{A\dot{A}} + i\,e\,A^{A\dot{A}})v_A + \frac{im}{\sqrt{2}}
\overline{u}^{\dot{A}} =0,\nonumber
\end{eqnarray}
where $\partial^{A\dot{A}} \equiv \sigma^{\alpha A\dot{A}}\partial_\alpha$, and 
$A^{A\dot{A}}=\sigma^{\alpha A\dot{A}}A_\alpha$; here 
$\sigma^{\alpha A\dot{A}}$ are the Infeld-van der Waerden symbols.

\noindent The electromagnetic potential is (see \cite{CR:pap1} for details),
\begin{equation}
 A^{A\dot{A}}=\frac{i}{e(u^cv_c)}\left\{
v^A\partial ^{B\dot{A}}
u_B+u^A\partial^{B\dot{A}}v_B+\frac{im}{\sqrt{2}}(u^A\overline{u}^{\dot{A}}+v^A
\overline{v}^{\dot{A}})\right\}.\label{emp:2sp}
\end{equation}
The reality conditions are,
\begin{eqnarray}
&&\partial^{A\dot{A}}(u_A\overline{u}_{\dot{A}}) = -
\frac{im}{\sqrt{2}}(u^C v_C -\overline{u}^{\dot{C}}\overline{v}_{\dot{C}})
\label{realc:2sp}\\
&&\partial^{A\dot{A}}(v_A\overline{v}_A)= \frac{im}{\sqrt{2}} (u^C v_C -
\overline{u}^{\dot{C}}\overline{v}_{\dot{C}})\nonumber\\
&&u_A\partial^{A\dot{A}}\bar{v}_{\dot{A}} -
\overline{v}_{\dot{A}}\partial^{A\dot{A}} u_A=0.\nonumber
     \end{eqnarray}
The Maxwell equations are,
\begin{equation}
\partial ^{\alpha}F_{\alpha \beta}
=-4\pi e\,j_\beta =
-4\pi e\,\sqrt{2}\sigma_\beta^{A\dot{A}}(u_A\overline{u}_{\dot{A}}+v_A\overline{v}_{\dot
{A}}).\label{max:2sp}
\end{equation}

The equations (\ref{emp:2sp}), (\ref{realc:2sp}) and (\ref{max:2sp}) 
are entirely equivalent  to the original Maxwell-Dirac equations, 
(\ref{dir:norm}) and (\ref{max:norm}).

\section{Isolated, Stationary Maxwell-Dirac Systems}

We recall the definitions of \cite{CH:pap3} for stationary and isolated systems.
\begin{definition}
 A Maxwell-Dirac system is said to be {\em stationary} 
if there is a gauge in which $\psi  =e^{i\omega t}\phi$, with the bi-spinor
$\phi$ independent of $t$. Such a gauge will be referred to as a stationary 
gauge.\label{def:st}\end{definition}

Clearly, a stationary gauge is not unique -- any gauge transformation 
$\psi\to e^{i\omega t}\psi$ leaves the system in a stationary gauge. Note that 
under such a gauge change 
we have, $A^{\alpha}\to A^{\alpha}-\frac{\omega}{e}\delta^{\alpha}_0$. We are 
interested in isolated systems, i.e. systems for which the fields decay 
suitably as $|x|\to \infty$, in 
this case we will require that $A^{\alpha}\to 0$ as $|x|\to\infty$ in some 
stationary gauge. In this particular gauge we will write $\psi =e^{-iEt}\phi$, 
for the stationary gauge in which $A^0\to 0$ as $|x|\to \infty$. Notice that 
for any stationary system in a stationary gauge $A^{\alpha}$ is independent 
of time, $t$, (see equation (\ref{emp:2sp})).

In most physical processes that we would wish to model using the Maxwell-Dirac
 system we would be interested in isolated systems -- systems where the fields
and sources are largely confined to a compact region of $\mathbb{R}^3$. This
requires that the fields decay sufficiently quickly as $|x|\to\infty$.

The best language for the discussion of such decay conditions and other
regularity issues is the language of
weighted function spaces; specifically weighted classical and Sobolev spaces.
In \cite{CH:pap3} the weighted Sobolev spaces, $W^{k,p}_\delta$, were used 
following the definitions of
\cite{RB:mass}. These definitions have the advantage that the decay
rate is explicit: under appropriate circumstances a function in $W^{k,p}_\delta$
 behaves as $|x|^\delta$ for large $|x|$. An element, $f$, of
$W^{k,p}_\delta$
has $\sigma^{-\delta+|\alpha|-\frac{3}{p}}\partial^{|\alpha |}f$ in $L^p$ for
each multi-index $\alpha$ for which $0\leq |\alpha|\leq k$; here $\sigma =
\sqrt{1+|x|^2}$ and we are working on $\mathbb{R}^3$ (or some appropriate
subset thereof) -- see \cite{RB:mass} or \cite{CBC:sob} and \cite{AGG:sob}
(the later papers use a different indexing of the Sobolev spaces). We will 
make use of the Sobolev inequality and frequent use of the
 multiplication lemma.

\noindent{\bf Sobolev Inequality} (see \cite{RB:mass}, \cite{CBC:sob}) 
{\em If $f\in W^{k,p}_\delta$ then
\[ \begin{array}{ll}
\mbox{(i)}&\|f\|_{\frac{np}{(n-kp)},\delta}\leq C\|f\|_{k,q,\delta},\ \ \mbox{if}\ \ n-kp>0\ \ \mbox{and}\ \ 
                                                                                 p\leq q\leq\frac{np}{(n-p)}\\
\mbox{(ii)}&\|f\|_{\infty ,\delta}\leq C\|f\|_{k,p,\delta},\ \ \mbox{if}\ \ n-kp<0,\ \ \mbox{and}\\
\, & |f(x)| = o(r^\delta )\ \ \mbox{as}\ \ r\to\infty.\end{array}\] }

\noindent{\bf Multiplication Lemma} (see \cite{CBC:sob})
 {\em Pointwise multiplication on $E_\rho$ 
is a continuous bilinear mapping
\[ W^{k_1,2}_{\delta_1}\times W^{k_2,2}_{\delta_2}\to W^{k,2}_\delta ,\]
if $k_1,k_2\geq k,\, \, \, k<k_1+k_2-n/2$, and $\delta >\delta_1+\delta_2$. }

We will be interested in the asymptotic region (spatially) of the Maxwell-Dirac
system, which we denote by $E_\rho =\mathbb{R}^3\backslash B_\rho$, where
$B_\rho$ is the ball of radius $\rho$. We will take our fields to be elements of
the function spaces $W^{k,p}_{\delta}(E_{\rho})$ for certain values of the 
indices $k$, $p$ and $\delta$. Before introducing the precise definition of an 
isolated system we must (following \cite{CH:pap3}) introduce some notation.

Suppose we have a stationary system and we are in a stationary gauge for which
$A^\alpha\to 0$ as $|x|\to\infty$. Write, $u_A=e^{-iEt}U_A$ and
$\bar{v}^{\dot{A}}=e^{-iEt}\bar{V}^{\dot{A}}$ with $U_A$, $V_A$ and $A^\alpha$
all independent of time, $t$. Note that $u_Cv^C=U_CV^C$ is a gauge and 
Lorentz invariant complex scalar function,
this means we can introduce a (unique up to sign) ``spinor dyad"
$\{ o_A, \iota_B\}$ with $\iota^Ao_A =1$ -- some facts on 2-spinor dyads are 
collected in Appendix A at the end of the paper. 
The dyad is defined as follows, let 
$U_CV^C =Re^{i\chi}$ -- where $R$ and $\chi$ are real functions -- then write,
\[ U_A =\sqrt{R}e^{i\frac{\chi}{2}}o_A\,\,\,\mbox{and}\,\,\, 
V_A=\sqrt{R}e^{i\frac{\chi}{2}}\iota_A.\]
Note that we must have $R>0$ (almost everywhere) because of our non-degeneracy 
condition. 

We can now define our isolated systems, note this definition is a little more 
general than the definition of \cite{CH:pap3}.
\begin{definition} A stationary Maxwell-Dirac system will be said to be isolated if, in some stationary gauge, we have
\[\psi = e^{-iEt}\sqrt{R}\left(\begin{array}{c}
                                      e^{\frac{i\chi}{2}}o_A\\
                                      e^{-\frac{i\chi}{2}}\bar{\iota}^{\dot{A}}
                       \end{array}\right),\]
with $E$ constant and 
$\sqrt{R}\in W^{3,2}_{-\tau}(E_\rho );\,\,\, e^{\frac{i}{2}\chi}o_A,
\,\,\, e^{\frac{i}{2}\chi}\iota_A
\in W^{3,2}_\epsilon (E_\rho )$ 
and\newline $A^\alpha\in W^{2,2}_{-1+\epsilon}(E_\rho )$, 
for some $\tau >\frac{3}{2}$, $\rho >0$ 
and any $\epsilon >0$.\label{def:iso}
\end{definition}
{\em Remarks}
\begin{itemize}
\item  This definition ensures, after use of the Sobolev inequality and the 
multiplication lemma, that $\psi =o(r^{-\tau +\epsilon})$ and $A^\alpha =
o(r^{-1+\epsilon})$.
\item Notice our condition places regularity restrictions on the fields in the 
region $E_{\rho}$ only. In the interior of $B_{\rho}$ there are no
regularity assumptions.
\item A minimal condition that one may impose
on the Dirac field is that it have finite total charge in the region $E_\rho$,
this amounts to\begin{eqnarray}
 \int_{E_{\rho}}\, j^0\, dx&=&\int_{E_{\rho}}\left( |U_0|^2+|U_1|^2+|V^0|^2+
|V^1|^2\right)\,dx\nonumber\\
  &=&\int_{E_{\rho}}R\left(|o_0|^2+|o_1|^2+|\iota^0|^2+
|\iota^1|^2\right)\,dx\ <\infty .\nonumber
\end{eqnarray} 
This, of course, simply means that $U_A$ and $V^A$ are in 
$L^2(E_\rho )$. So $U_A$ and $V^A$ would have $L^2$ decay at infinity; 
roughly, they would decay faster than $|x|^{-\frac{3}{2}}$, i.e. we require
at least $\tau >\frac{3}{2}$.
\item 
The spherically symmetric solution of \cite{CR:pap1} provides an 
excellent example of an {\em isolated}, {\em
stationary} and {\em static} Maxwell-Dirac system.
\end{itemize}

With our assumption that the Maxwell-Dirac system is isolated and stationary we 
can impose the Lorenz gauge condition, without altering the 
above regularity and decay assumptions on $A^\alpha$. To see this we note that 
\[\partial_{\alpha}A^{\alpha}=\sum^{3}_{j=1}\partial_j A^j\in 
W^{1,2}_{-2+\epsilon}(E_\rho )\]
 and that the Laplacian, $\Delta$, gives an 
isomorphism $W^{3,2}_{\epsilon}\to W^{1,2}_{-2+\epsilon}$ (we may assume 
$\epsilon <1$) -- see \cite{NW:Ell}, \cite{Mc:Lap} (and also \cite{RB:mass}, 
\cite{CBC:sob} and \cite{AGG:sob}). This 
means there is a unique solution, $\Omega\in W^{3,2}_{\epsilon}(E_\rho )$, of 
the equation
\[ \Delta\Omega +\partial_\alpha A^\alpha =0.\]
Consequently, for the gauge change $A_\alpha\to \hat{A}_{\alpha}=A_\alpha +
\partial_\alpha\Omega$ we still have $\hat{A}^{\alpha}\in W^{2,2}_{-1+\epsilon}$
, with $\hat{A}^\alpha$ satisfying the Lorenz gauge condition. 

The electromagnetic potential of our stationary, isolated Maxwell-Dirac system 
will be taken to satisfy the two equations,
\begin{eqnarray}
\Delta A^{\alpha}&=&4\pi e\sqrt{2}\sigma^{\alpha A\dot{A}}R\left( 
o_A\bar{o}_{\dot{A}} +\iota_A\iota_{\dot{A}}\right)\label{max:is}\\
\partial_{\alpha} A^{\alpha}&=&\sum^3_{j=1}\frac{\partial A^j}{\partial x^j}=0.\label{lorz:is}
\end{eqnarray}

To end this section we present a simple result which we will need in the 
following sections.
\begin{lemma} For a stationary and isolated Maxwell-Dirac system, 
in the Lorenz gauge,
\begin{eqnarray}
&&A^0 - \frac{q_0}{|x|}\in W^{5,2}_{-\eta}(E_\rho ),\,\,\, q_0\,\,\,
\mbox{a constant}
\,\,\,\mbox{and}\\
&&A^j \in W^{5,2}_{-\eta}(E_\rho ) ,\,\,\, j=1,2,3;\,\,\,\mbox{and}\,\,\,\eta 
=2(\tau -1)>1. 
\end{eqnarray}\label{lem:A}\end{lemma}
{\em Proof}. 
Firstly we note that the source term of the Maxwell equation 
(\ref{max:is}) is in $W^{3,2}_{-2\tau +2\epsilon}(E_\rho)$, i.e. $j^{\alpha}\in 
W^{3,2}_{-2\tau +2\epsilon}(E_\rho)$. The Laplacian gives an isomorphism between
 $W^{5,2}_{-2(\tau -\epsilon -1)}$ and $W^{5,2}_{-2\tau +2\epsilon}$ see 
\cite{NW:Ell}, \cite{Mc:Lap} (also \cite{RB:mass}, 
\cite{CBC:sob} and \cite{AGG:sob}). So there exists an 
$a^\alpha\in W^{5,2}_{-2(\tau -\epsilon -1)}=W^{5,2}_{-\eta}$ such that,
\[ \Delta a^{\alpha} = 4\pi ej^{\alpha}.\]
Now, as $A^{\alpha}\in W^{2,2}_{-1+\epsilon}(E_\rho )$, we have 
\[ A^{\alpha}=
\frac{q_{\alpha}}{|x|}+ a^{\alpha},\]
where the $q_{\alpha}$ are constants. 
Applying the Lorenz condition we find $q_{j}= 0$, for $j=1,2,3$.
\hfill{$\Box$}

We can improve the decay rates here to higher (negative) order harmonic 
polynomials, at the expense of regularity, by using the Dirac equations to get 
$j^{\alpha}\in W^{2,2}_{-2(\tau +1)+2\epsilon}(E_\rho)$ -- but 
lemma \ref{lem:A} is sufficient for our purposes.

The constant $q_0$ is the total electric charge of the system (i.e. the 
electric charge of the Dirac field plus the charge due to any external sources 
in $B_\rho$); this is easily 
seen by taking a Gauss 
integral over the sphere at infinity of the electrostatic field (given by the 
gradient of $A^0$).

\section{No Embedded Eigenvalues}

A famous theorem of H. Weyl asserts the invariance of the essential spectrum of 
the perturbation of an operator if the difference of the resolvents of 
the perturbed and original 
operators is compact, see \S{4.3.4} of \cite{BT:dbook}. In standard notation we 
have, for a stationary system with $\psi =e^{-iEt}\phi$,
\[ H\phi = E\phi ,\,\,\,\mbox{with}\,\,\, H=\gamma^0\sum^{3}_{j=1}\gamma^j
\left(-i\frac{\partial}{\partial x^j}+eA^j\right)
+\left(\gamma^0 m -e A^0\right) ,\]
the free operator, $H_0$, has $A^\alpha =0$. Consequently, we have
\begin{proposition} The Dirac Hamiltonian operator $H$ of a stationary and
 isolated Maxwell-Dirac system has the same essential spectrum as the free 
operator, i.e.
\begin{eqnarray}
\sigma_{\mbox{\small ess}}(H)&=&\sigma_{\mbox{\small ess}}(H_0)\nonumber\\
                        &=&\sigma (H_0)\nonumber\\
                        &=&(-\infty ,-m]\cup [m,\infty ).\nonumber
\end{eqnarray}
\end{proposition}
{\em Proof}. A simple adaption of \S{4.3.4} of \cite{BT:dbook}.
\hfill{$\Box$}

We will now prove that for our stationary isolated systems there are no embedded
 eigenvalues, $E$ (the ``energy").
\begin{theorem} A stationary and isolated Maxwell-Dirac system has no embedded 
eigenvalues $E$, i.e. $\displaystyle{-m\leq E\leq m}$. In particular, the 
following limit exists,
\begin{eqnarray}
\frac{E}{m}&=&\lim_{|x|\to\infty}\frac{\cos\chi}
{\sqrt{1+\frac{1}{2}\lambda^2}},\,\,\,
\mbox{where}\,\,\, \lambda^2=\sum^3_{j=1}\left( l^j+n^j\right)^2,\nonumber\\
&&\mbox{with}\,\,\, l^{\alpha} = \sigma^{\alpha}_{A\dot{A}}o^A\bar{o}^{\dot{A}},
\,\,\,\mbox{and}\,\,\, n^{\alpha}= \sigma^{\alpha}_{A\dot{A}}\iota^A
\bar{\iota}^{\dot{A}}.\nonumber
\end{eqnarray}
\end{theorem}
{\em Remarks}
\begin{itemize}
\item The result needs the Maxwell equations only in order to derive the decay 
result for $A^\alpha$ of lemma \ref{lem:A}. The result remains true for the 
Dirac equation alone if we assume the appropriate decay for $A^\alpha$.
\item The only conditions required are the rather weak regularity and decay 
conditions of an isolated system. No positivity conditions on the potential 
are required, {\em cf} \S{4.7.2} of \cite{BT:dbook}.
\end{itemize}

{\em Proof}. The proof is remarkably simple, it is a matter of 
exploiting the notation introduced in definition \ref{def:iso}. We begin by 
re-writing the 
2-spinor form of the Dirac equations, (\ref{dir:2sp}), in this notation. We 
have,
\begin{eqnarray} \frac{o_A}{2}\left(\frac{\partial^{A\dot{A}}R}{R}+
i\partial^{A\dot{A}}\chi\right)+\partial^{A\dot{A}}o_A -ieB^{A\dot{A}}
o_A +\frac{im}{\sqrt{2}}e^{-i\chi}\,\bar{\iota}^{\dot{A}}&=&0\label{dir:rchi}\\
\frac{\iota_A}{2}\left(\frac{\partial^{A\dot{A}}R}{R}+
i\partial^{A\dot{A}}\chi\right)+\partial^{A\dot{A}}\iota_A +ieB^{A\dot{A}}
\iota_A +\frac{im}{\sqrt{2}}e^{-i\chi}\,\bar{o}^{\dot{A}}&=&0;\nonumber
\end{eqnarray}
where, $B^\alpha = E\delta^{\alpha}_0+A^\alpha$.

We combine these two equations into a single (equivalent) equation which gives
the derivative $\frac{\partial_{\alpha}R}{R}+i\partial_{\alpha}\chi$. To do this
 multiply the first equation by $\iota_B$ and the second by $o_B$ and subtract.
 Using $o_A\iota_B-o_B\iota_A =\epsilon_{AB}$ (see Appendix A), we have
\begin{eqnarray}
&&\frac{\partial_{A\dot{A}}R}{R}+i\partial_{A\dot{A}}\chi-2i\left(\iota_Ao_B+
\iota_Bo_A\right)B^{B}_{\,\,\dot{A}}\label{rchi:eq}\\&&+2\left(\iota_A
\partial^{B}_{\,\,\dot{A}}o_B
-o_A\partial^B_{\,\,\dot{A}}\iota_B\right)
+\sqrt{2}im\left(\iota_A
\bar{\iota}_{\dot{A}}-
o_A\bar{o}_{\dot{A}}\right) e^{-i\chi}=0.\nonumber
\end{eqnarray}
We now use the multiplication lemma 
to place the terms of the 
equation into an appropriate weighted Sobolev space:
\begin{itemize}
\item $\partial_{A\dot{A}}\chi\in W^{2,2}_{-1+2\epsilon}(E_\rho )$, since 
(from definition \ref{def:iso}),
\[ e^{i\chi}= e^{i\frac{\chi}{2}}\,\iota^Ae^{i\frac{\chi}{2}}\,o_A\in 
W^{3,2}_{2\epsilon}(E_\rho );\]
\item $(\iota_Ao_B+\iota_Bo_A)B^B_{\,\,\dot{A}}=(\iota_Ao_B+\iota_Bo_A)
(E\sigma^{0\,B}_{\phantom{0}\,\,\,\dot{A}}+A^B_{\,\,\dot{A}})$, 
breaking this into two terms we have,
\[ (\iota_Ao_B+\iota_Bo_A)E\sigma^{0\,B}_{\phantom{0}\,\,\,\dot{A}}\in 
W^{3,2}_{2\epsilon}(E_\rho ),\,\,\, (\iota_Ao_B+\iota_Bo_A)A^B_{\,\,\dot{A}}\in 
W^{2,2}_{-1+2\epsilon}(E_\rho ),\]
this last inclusion uses lemma \ref{lem:A};
\item $\iota_A\partial^{B}_{\,\,\dot{A}}o_B
-o_A\partial^B_{\,\,\dot{A}}\iota_B\in W^{2,2}_{-1+2\epsilon}(E_\rho ).$
\end{itemize}
So that equation (\ref{rchi:eq}) implies,
\begin{eqnarray} &&\frac{\partial_{A\dot{A}}R}{R}-2i(\iota_Ao_B +\iota_Bo_A)E
\sigma^{0\,B}_{\phantom{0}\,\,\,\dot{A}}\label{Rchi:inW}\\
&&+\sqrt{2}im\left(\iota_A\iota_{\dot{A}}-
o_A\bar{o}_{\dot{A}}\right) e^{-i\chi}\in W^{2,2}_{-1+2\epsilon}(E_\rho ).
\nonumber\end{eqnarray}
We now contract equation (\ref{Rchi:inW}) with $o^A\bar{o}^{\dot{A}}$ and so on.
 Using, the results of Appendix A (in particular the last two facts), 
we have, after splitting the resulting equations into real and imaginary 
parts, 
\begin{eqnarray}&&\frac{\partial_\alpha R}{R}+\sqrt{2}m\sin\chi\,
(n_{\alpha}-l_{\alpha})\label{dR:inW}\\
&&+i(m^0-\bar{m}^0)E(m^0\bar{m}_{\alpha}-\bar{m}^om_{\alpha})\in 
W^{2,2}_{-1+4\epsilon}(E_\rho ),\nonumber\\
&&(l^0-n^0),\, (m^0+\bar{m}^0)E,\,(l^0+n^0)E-\sqrt{2}m\cos\chi\,\in
W^{2,2}_{-1+4\epsilon}(E_\rho ).\label{cschi:inW}
\end{eqnarray}
Now (using Appendix A), 
\[ l^0+n^0 =\frac{1}{\sqrt{2}}\left( |o_0|^2+|o_1|^2+
|\iota_0|^2+|\iota_1|^2\right) >0\,\,\,\mbox{and}\]
\[( l^\alpha +n^\alpha )\eta_{\alpha\beta}
( l^\beta +n^\beta ) =
( l^0+n^0)^2 -\sum^3_{j=1}(l^j+n^j)^2 = 2,\]
 here $\eta_{\alpha\beta}$ is the 
Minkowski metric. So we have,
\[ \frac{1}{\sqrt{2}}(l^0+n^0) =\frac{1}{\sqrt{2}}(l_0+n_0) = 
\sqrt{1+\frac{1}{2}\sum^3_{j=1}(l^j+n^j)^2}=\sqrt{1+\frac{1}{2}
\lambda^2}.\]
From the last inclusion of equation (\ref{cschi:inW}) we have
\[\frac{1}{\sqrt{2}}(l^0+n^0)E-m\cos\chi\, =\sqrt{1+\frac{1}{2}\lambda^2}\,\, E
-m\cos\chi\,\in W^{2,2}_{-1+4\epsilon}(E_\rho ).\]
Notice that, $\displaystyle{(1+\frac{1}{2}\lambda^2)^{-\frac{1}{2}}\in 
W^{3,2}_{2\epsilon}}$ and consequently,
\[\frac{E}{m}-\frac{\cos\chi}{\sqrt{1+\frac{1}{2}\lambda^2}}\in 
W^{2,2}_{-1+6\epsilon}(E_\rho ).\]
Finally, from the Sobolev inequality we have,
\[ \left|\frac{E}{m}-\frac{\cos\chi}{\sqrt{1+\frac{1}{2}\lambda^2}}\right|< C\,
|x|^{-1+6\epsilon},\]
for any $\epsilon >0$ and some constant $C$. Hence the limit $|x|\to \infty$ of 
the left hand side 
exists and is zero, which completes the proof.
\hfill{$\Box$}

\section{Regularity and Decay}

A stationary Maxwell-Dirac system is an elliptic system of partial differential
 equations. So it should be a simple matter to apply the theory of elliptic 
regularity to obtain the best possible regularity results for the Maxwell and 
Dirac fields. That this is indeed the case is demonstrated in the next theorem.
\begin{theorem}A stationary Maxwell-Dirac system for which,
$U_A,\,\,\, V_A\in L^2(E_\rho )$ and $A^{\alpha}\in
L^1_{\mbox{\small{loc}}}(E_\rho )$, 
is $C^{\infty}$, i.e. $U$, $V$ and $A$ are in $C^{\infty}(E_\rho )$.
\end{theorem}
{\em Proof}.
First we note that we can
always find a gauge transformation which takes $A$ into the Lorenz gauge
while leaving it in the same Lebesgue space.

We have a set of Poisson equations for the $A$'s (equation (\ref{max:is})),
and the elliptic Klein-Gordon equations (B.3). These equations take the form,
\begin{eqnarray}
&&\Delta A^{\alpha}=4\pi ej^{\alpha}\nonumber\\
&&\Delta U_A+2ie\sum^3_{j=1}A^j\partial_jU_A= (m^2-E^2)U_A +h^B_AU_B,\nonumber
\end{eqnarray}
where $h^B_A$ is a quadratic function of the $A$'s; with a similar equation
for the $V$'s.
As $U$ and $V$ are in $L^2(E\rho )$, $j^\alpha$ is in $L^1(E_\rho)$, so the
$A$'s must be in
$L^3_{\mbox{\small{\em loc}}}(E_\rho )$. Putting this information into the $U$
 and $V$ equations we conclude (after the use of the H\"{o}lder inequality)
 that the $h^A_BU_A$ are in $L^{\frac{6}{5}}_{\mbox{\small{\em loc}}}(E_\rho )$ 
and so the $U$'s and $V$'s are in $L^{3}_{\mbox{\small{\em loc}}}
(E_\rho )$ (elliptic regularity).
 We can now conclude 
that all the fields are in $C^{0,\alpha}$ and then in $C^{2,\alpha}$. 
Iterating, we finally find that
the fields are in $C^{\infty}(E_\rho )$ (this is the classical `bootstrap' 
argument, see chapter 10 of \cite{LL:book} or
chapter 9 of \cite{GT:book}).\hfill{$\Box$}

In the case of the free Klein-Gordon equation it is easy to see that, in the 
stationary case, if $E^2-m^2<0$ then the Dirac field decays exponentially. 
This fact remains true for our stationary isolated systems.
\begin{theorem} For an isolated and stationary Maxwell-Dirac system, with
$E^2\neq m^2$, the Dirac 
fields $U$ and $V$ (along with all their derivatives) decay exponentially as 
$|x|\to\infty$.
\end{theorem} 
{\em Proof}.
With $E^2\neq m^2$ and theorem 1 we have $|E|<m$.

We now cast our Klein-Gordan equations in a form suitable for the application 
of the maximum principle. 
Firstly, $\Delta (|U|^2)= \bar{U}\Delta U+U\Delta\bar{U}+2|\nabla U|^2$, so
using equation (B.3):
\begin{eqnarray}
&\Delta (|U_0|^2+|U_1|^2)&\label{thm3:eq}\\&=&\nonumber\\
&2ie\sum^{3}_{j=1}(U_0A^j\partial_j\bar{U}_0-\bar{U}_0A^j\partial_jU_0
U_1A^j\partial_j\bar{U}_1-\bar{U}_1A^j\partial_jU_1)&\nonumber\\
&+2(m^2-E^2)(|U_0|^2+|U_1|^2)
+2(|\nabla U_0|^2+|\nabla U_1|^2)&\nonumber\\
&-2\left[ 2eEA^0+e^2A^{\alpha}A_{\alpha}\right] (|U_0|^2+|U_1|^2)
-ie(\bar{\partial}A-\partial\bar{A})(|U_0|^2-|U_1|^2)&\nonumber\\
&-2ie\left[(-\partial_z A+\partial A^3) U_0\bar{U_1}-
(-\partial_z \bar{A}+\bar{\partial} A^3)\bar{U_0}U_1\right]&\nonumber
\end{eqnarray}
We will use the following comparison function,
\[ w(x)=C_0\frac{e^{-\sqrt{2}k|x|}}{|x|}\,\,\,\mbox{with}\,\,\,  
C_0=\rho e^{\sqrt{2}k\rho}\sup_{|x|=\rho}[h(x)]\,\,\,\mbox{and}\,\,\,
k>0,\]
where $h(x)=(|U_0|^2+|U_1|^2+|V_0|^2+|V_1|^2)$; the supremum in the
definition of $C_0$ is well-defined and finite since we have by the Sobolev
inequality, $0<h(x)< C|x|^{-2(\tau-\epsilon)}$, on $E_\rho$, for
$\rho$ large enough. Note that we have $\Delta w-2k^2w=0$, on $E_\rho$.

We show that
\[ \Delta [h(x)-w(x)]-2k^2[h(x)-w(x)]\geq 0.\]
The following inequalities will be needed -- 
 in each case we have used the Sobolev inequality on 
the $A$'s (after use of lemma 1). Firstly,
\begin{eqnarray}i\sum^{3}_{j=1}&&\hspace{-0.8cm}(U_0A^j\partial_j\bar{U}_0-\bar{U}_0A^j
\partial_jU_0
U_1A^j\partial_j\bar{U}_1-\bar{U}_1A^j\partial_jU_1)\nonumber\\
&=&R\sum^{3}_{j=1}\left[ A^j\partial_j\chi (|o_0|^2+|o_1|^2)\right.\nonumber\\
&&\left.\phantom{= R \sum}+i(o_0A^j\partial_j\bar{o}_0-\bar{o}_0A^j\partial_jo_0
+o_1A^j\partial_j\bar{o}_1-\bar{o}_1A^j\partial_jo_1)\right] \nonumber\\
&>&-2\frac{C_1}{|x|}R\nonumber\\
&>&-\frac{C_1}{|x|}R(|o_0|^2+|o_1|^2+|\iota_0|^2+|\iota_1|^2)\nonumber\\
&=&-\frac{C_1}{|x|}(|U_0|^2+|U_1|^2+|V_0|^2+|V_1|^2),\nonumber\end{eqnarray}
where we have used,
\[ 1=|\iota^Ao_A|=|\iota^0o_0+\iota^1o_1|\leq 
\frac{1}{2}(|o_0|^2+|o_1|^2+|\iota_0|^2+|\iota_1|^2).\]

Next, in a similar vein, we have 
\begin{eqnarray}
-2\left[ 2eEA^0+e^2A^{\alpha}A_{\alpha}\right] &>&-\frac{C_2}{|x|};
\nonumber\\&&\nonumber\\
-ie(\bar{\partial}A-\partial\bar{A})(|U_0|^2-|U_1|^2)&>&
-e|\bar{\partial}A-\partial\bar{A}|(|U_0|^2+|U_1|^2)\nonumber\\
&>&-\frac{C_3}{|x|^2}(|U_0|^2+|U_1|^2);\nonumber\\&\nonumber\end{eqnarray}
\begin{eqnarray}
-2ie\left[(-\partial_z A+\partial A^3) U_0\bar{U_1}\right.\hspace{-0.5cm}&&
\hspace{-0.5cm}\left.-(-\partial_z \bar{A}+\bar{\partial} A^3)\bar{U_0}U_1
\right]
\nonumber\\
&>&-2\frac{C_4}{|x|^2}|U_0U_1|\nonumber\\
&>&-\frac{C_4}{|x|^2}(|U_0|^2+|U_1|^2);\nonumber
\end{eqnarray}
here the $C_j$ are positive constants.

With the use of equation (\ref{thm3:eq}) and the inequalities we have,
\begin{eqnarray}
 &\Delta [(|U_0|^2+|U_1|^2)]&\nonumber\\
&\geq&\nonumber\\
&\left[ 2(m^2-E^2)-\frac{C_2}{|x|}-\frac{C_3}{|x|^2}-\frac{C_4}{|x|^2}-
\right] 
(|U_0|^2+|U_1|^2)&\nonumber\\&-\frac{C_1}{|x|}h(x)+2(|\nabla U_0|^2+
|\nabla U_1|^2).&\nonumber
\end{eqnarray}
There is, of course, an entirely similar equation for $|V^0|^2+|V^1|^2$.
Adding these two equations gives,
for every $k$ such that $0<k<\sqrt{m^2-E^2}$,
\begin{eqnarray}&\Delta\hspace{-.3cm}&[h(x)-w(x)]-2k^2[h(x)-w(x)]\nonumber \\
&&\geq\left[ 2(m^2-E^2-k^2)-\frac{C}{|x|}\right]h(x) +(|\nabla U_0|^2+|\nabla U_1|^2
+|\nabla V_0|^2 +|\nabla V_1|^2)\nonumber\\
&&\geq 0,\,\,\,\mbox{for }\rho\,\,\,\mbox{large enough}.\nonumber\end{eqnarray}
Applying the maximum principle on $E_\rho$ we see that the non-negative maximum 
of $h(x)-w(x)$ must occur at infinity or on $|x|=\rho$. However,
\begin{eqnarray}
\lim_{|x|\to\infty}\left[ h(x)-w(x)\right]=0&&\,\,\,
\mbox{and}\nonumber\\
\left[ h(x)-w(x)\right]_{|x|=\rho}\leq 0.&&\nonumber
\end{eqnarray}
We conclude that $\left[ h(x)-w(x)\right] \leq 0$ on $E_\rho$, so 
that
\[ |U_0|^2+|U_1|^2+|V^0|^2+|V^1|^2\leq C_0\frac{e^{-\sqrt{2}k|x|}}{|x|}.\]
So the $U$'s and $V$'s decay exponentially.

Differentiating the Klein-Gordon equations we can use the same procedure to 
show that the first derivatives decay exponentially. After taking account of 
theorem 2 we can iterate this procedure once we note that the solution for 
$A^\alpha$ can be written as the sum of an harmonic polynomial (of negative 
degree) and the convolution of $j^\alpha$ and the appropriate Green's function.
\hfill{$\Box$}

This theorem does not deal with decay of solutions in the case 
$|E|=m$. It is 
clear that the solutions in this case need not decay exponentially. 
The spherically symmetric 
solution of \cite{CR:pap1} is a case in point, this solution has $|E|=m$ and 
$R\sim\frac{C_0}{|x|^4}$ as $|x|\to\infty$.

The case $|E|=m$ is in a sense quite unique, as we will see in the next 
section.

\section{$|E|=m$ and Asymptotically Static Systems}

In this section we will prove a theorem which neatly ties together 
the $|E|=m$ case and the concept of an asymptotically static solution.

In \cite{CH:pap3} the idea of a static Maxwell-Dirac system was exploited to 
show that if the system was also stationary and isolated then the system was 
necessarily electrically neutral, with $|E|=m$. A static Maxwell-Dirac system 
is one for which
 (in some Lorentz frame) the spatial components of the Dirac current vanish. 
With the Dirac current written as,
\begin{eqnarray}
j^{\alpha}&=&\sqrt{2}\sigma^{\alpha}_{\phantom{\alpha}A\dot{A}}(u^A\bar{u}^{\dot{A}}+
v^A\bar{v}^{\dot{A}})\nonumber\\
&=&\sqrt{2}R(l^{\alpha}+n^{\alpha})\nonumber\\
&=&\sqrt{2}R\sigma^{\alpha}_{\phantom{\alpha}A\dot{A}}(o^A\bar{o}^{\dot{A}}
+\iota^A\bar{\iota}^{\dot{A}}),\nonumber
\end{eqnarray}
we would require for a static system that 
$l^k+n^k=0$, $k=1,2,3$.
We will now generalise this concept to that of an {\em asymptotically static} 
Maxwell-Dirac system.
\begin{definition} A Maxwell-Dirac system will be called asymptotically static, 
with decay rate $\kappa$ and differentiability index $s$,  
if $l^k+n^k=\sigma^{k}_{\phantom{k}A\dot{A}}(o^A\bar{o}^{\dot{A}}
+\iota^A\bar{\iota}^{\dot{A}}) \in W^{s,2}_{-\kappa}(E_\rho )$
for some $\rho$, $\kappa >0$ and $k=1,2,3$.\label{def:asyst}
\end{definition}
So an asymptotically static system decays toward a static system, as 
$|x|\to\infty$. The unit vector, $\frac{1}{\sqrt{2}}(l^\alpha +n^\alpha )$
 (in the direction of the current, $j^\alpha$) has only a time-like component 
in the limit as $|x|\to\infty$.

To take full advantage of this definition we will need to recast it in terms of 
the individual variables $o_A$, and $\iota_A$, this is done in the following 
lemma. 
\begin{lemma} An isolated and stationary 
Maxwell-Dirac system is asymptotically static if and only if  
\[ \iota^0 -\bar{o}_{\dot{0}},\,\,\,\iota^1-\bar{o}_{\dot{1}}\in 
W^{t,2}_{-\kappa +\epsilon}(E_\rho ).\]
where $t=min[3,s]$.\label{lem:asyst1}
\end{lemma}
{\em Proof}. First assume the system is asymptotically static with decay rate 
$\kappa$ and differentiability $s$

From the proof of theorem 1 and definition \ref{def:asyst} we have,
\begin{eqnarray}
(|o_0|^2+|o_1|^2+|\iota^0|^2+|\iota^1|^2)^2-4&=&2(l^0+n^0)^2-4\nonumber\\
&=&2\sum^{3}_{k=1}(l^k+n^k)^2\in W^{s,2}_{-2\kappa}(E_\rho ).\nonumber
\end{eqnarray}
Consequently, as $ [(|o_0|^2+|o_1|^2+|\iota^0|^2+|\iota^1|^2)+2]^{-1}\in 
W^{3,2}_{2\epsilon}(E_\rho )$, we have
\[|o_0|^2+|o_1|^2+|\iota^0|^2+|\iota^1|^2 -2\in 
W^{t,2}_{-2\kappa +2\epsilon}(E_\rho ).\]
Writing out the $l^k+n^k$ explicitly we have, 
\begin{eqnarray}
-(o_0\bar{o}_{\dot{1}}+o_1\bar{o}_{\dot{0}})+(\iota^0\bar{\iota}^{\dot{1}}
+\iota^1\bar{\iota}^{\dot{0}})&\in&  W^{s,2}_{-\kappa}(E_\rho )\nonumber\\
-(-o_0\bar{o}_{\dot{1}}+o_1\bar{o}_{\dot{0}})+(-\iota^0\bar{\iota}^{\dot{1}}
+\iota^1\bar{\iota}^{\dot{0}})&\in & W^{s,2}_{-\kappa}(E_\rho )\nonumber\\
-|o_0|^2+|o_1|^2+|\iota^0|^2-|\iota^1|^2&\in& W^{s,2}_{-\kappa}(E_\rho ).
\nonumber\end{eqnarray}
So we conclude that,
\begin{eqnarray} &-o_0\bar{o}_{\dot{1}}
+\iota^1\bar{\iota}^{\dot{0}}\in  W^{s,2}_{-\kappa}(E_\rho )\,\,\,\,\mbox{and}&
\nonumber\\
&|o_0|^2+|\iota^1|^2-1,\,\,\, |o_1|^2+|\iota^0|^2-1\in 
W^{t,2}_{-\kappa}(E_\rho ).&\nonumber
\end{eqnarray}
Writing these equations as a single matrix equation, 
\[\left(\begin{array}{rr}o_1&-\iota^0\\o_0&\iota^1\end{array}\right)
  \left(\begin{array}{ll}\bar{o}_{\dot{0}}-\iota^0&\bar{o}_{\dot{1}}-\iota^1\\
         \bar{\iota}^{\dot{1}}-o_1&o_0-\bar{\iota}^{\dot{0}}
\end{array}\right)=
\left(\begin{array}{cc}o_1\bar{o}_{\dot{0}}-\iota^0\bar{\iota}^{\dot{1}}&
|o_1|^2+|\iota^0|^2-1\\
|o_0|^2+|\iota^1|^2-1& o_0\bar{o}_{\dot{1}}-\iota^1\bar{\iota}^{\dot{0}}
\end{array}\right)\]
is in $W^{t,2}_{-\kappa}(E_\rho )$. The first matrix on the left has determinant
 $1$ and inverse,
\[\left(\begin{array}{rr}o_1&-\iota^0\\o_0&\iota^1\end{array}\right)^{-1}=
\left(\begin{array}{rr}\iota^1&\iota^0\\-o_0&o_1\end{array}\right) ,\]
which is in $W^{3,2}_{\epsilon}(E_\rho )$. The result now follows from the 
multiplication lemma after applying this inverse matrix to the previous 
equation.

Next, assuming $\iota^0 -\bar{o}_{\dot{0}},\,\,\,\iota^1-\bar{o}_{\dot{1}}\in 
W^{t,2}_{-\kappa +\epsilon}(E_\rho )$, we easily find that $l^k+n^k\in W^{t,2}_
{-\kappa +2\epsilon}(E_\rho )$, $k=1,2,3$, and any $\epsilon >0$. So the system 
is asymptotically static.
\hfill{$\Box$}

Now to our theorem connecting the two apparently unrelated notions, 
the condition $|E|=m$ and the idea of an asymptotically static system.
\begin{theorem} A stationary and isolated Maxwell-Dirac system is asymptotically
 static if $|E|=m$.\label{thm:emst}
\end{theorem}
{\em Proof}. Assume the system is stationary and isolated with $|E|=m$; write 
$E=\varepsilon m$, with $\varepsilon =\pm 1$. 

The proof is very simple, it simply involves manipulating expressions 
obtained in the proof of theorem 1. From that proof we have,
\[ \sqrt{1+\frac{1}{2}\lambda^2}-\varepsilon\cos\chi\,\in 
W^{2,2}_{-1+4\epsilon}(E_\rho ).\]
Note that we must have 
$\varepsilon\cos\chi\,\geq 0$, for $\rho$ large enough, so 
\[ \sqrt{1+\frac{1}{2}\lambda^2}+\varepsilon\cos\chi\,\in 
W^{2,2}_{4\epsilon}(E_\rho ).\] 
Multiplying the last two expressions and using 
the multiplication lemma, we have
\[ \sin^2\chi\, +\frac{1}{2}\lambda^2\in W^{2,2}_{-1+8\epsilon}(E_\rho ).\] 
These equations together with the fact that both $\sin^2\chi$ and 
$l^k+n^k$ are in $W^{3,2}_{2\epsilon}(E_\rho )$ enable 
us to conclude that 
\[ \sin\chi\, ,\, l^k+n^k \in W^{2,2}_{-\frac{1}{2}+4\epsilon}(E_\rho ).\]
The system is, according to definition 3, asymptotically static.
\hfill{$\Box$}

\section{Exponential Decay, the $|E|=m$ Case}

In this section we will prove that the Dirac field decays exponentially in the 
$|E|=m$ case as well -- at least when the total charge $q_0\neq 0$. In fact we 
obtain tight bounds on the decay of the Dirac field in this case.
\begin{theorem} The Dirac field for a stationary, isolated and 
asymptotically static Maxwell- Dirac 
system, with $|E|=m$, $\kappa >1$,  
$s=3$ (definition 3) and $q_0\neq 0$ decays exponentially as $|x|\to\infty$. In 
fact, there exists two positive constants $C_1$ and $C_2$ such that
\[ C_1\frac{e^{-4\sqrt{2}m\lambda\sqrt{|x|}}}{|x|^{\frac{3}{2}}}<R<
 C_2\frac{e^{-4\sqrt{2}m\lambda\sqrt{|x|}}}{|x|^{\frac{3}{2}}},\]
where $\lambda >0$, $\lambda^2=-\varepsilon e\frac{q_0}{m}$ is necessarily 
positive and $\frac{E}{m}=\varepsilon=\pm 1$.
\label{thm:Emdecay}\end{theorem}
Before embarking on a proof of this theorem we will need a couple of preparatory
 lemmas. In the course of proving the second of these two lemmas we will 
also show incidentally that the electric dipole moment must vanish. 
We are assuming that $|E|=m$ and 
write $\frac{E}{m}=\varepsilon$. As $\cos\chi\to\varepsilon$, with $|x|\to\infty
$ we can take $\chi =n\pi +\zeta$ where $(-1)^n=\varepsilon$ and $\zeta\to 0$, 
as $|x|\to\infty$.

We require a more careful analysis of equation 
(\ref{rchi:eq}). The 
imaginary part of this equation is,
\begin{eqnarray}\partial_{\alpha}\chi &+2(E+eA^0)(n^0l_{\alpha}-l^0n_\alpha )+
2e\sum^{3}_{k=1}A^k(l^kn_{\alpha}-n^kl_{\alpha})\label{eq:chi}\\
&+\sqrt{2}m\cos\chi\, 
(n_\alpha -l_\alpha ) -i\left[ (\bar{o}_{\dot{A}}\partial^{A\dot{A}}o_A-
o_A\partial^{A\dot{A}}\bar{o}_{\dot{A}})n_\alpha\right.\nonumber\\ 
&+(\bar{\iota}^{\dot{A}}\partial_{A\dot{A}}\iota^A
-\iota^A\partial_{A\dot{A}}\bar{\iota}^{\dot{A}})l_\alpha
+(\bar{o}_{\dot{A}}\partial_{A}^{\phantom{A}\dot{A}}\iota^A
-\iota^A\partial_A^{\phantom{A}\dot{A}}\bar{o}_{\dot{A}})m_\alpha\nonumber\\
&+(\bar{\iota}^{\dot{A}}\partial^{A}_{\phantom{A}\dot{A}}o_A-\left.
o_A\partial^{A}_{\phantom{A}\dot{A}}\bar{\iota}^{\dot{A}})\bar{m}_\alpha\right] 
=0.\nonumber
\end{eqnarray}
Assuming the system is asymptotically static and making use of lemma 2 
simple calculation reveals,
\begin{eqnarray}
\bar{\iota}^{\dot{A}}\partial_{A\dot{A}}\iota^A+o_A\partial^{A\dot{A}}
\bar{o}_{\dot{A}}&\in W^{2,2}_{-\kappa-1+2\epsilon}(E_\rho )\nonumber\\
\iota^A\partial_A^{\phantom{A}\dot{A}}\bar{o}_{\dot{A}}-\bar{o}_{\dot{A}}
\partial_{A}^{\phantom{A}\dot{A}}\iota^A&\in 
W^{2,2}_{-\kappa-1+2\epsilon}(E_\rho ).
\nonumber
\end{eqnarray}
From lemma 2, $n^0-l^0\in W^{3,2}_{-\kappa+2\epsilon}(E_\rho )$.

For $\alpha =k=1,2,3$ we make use of 
$n^k+l^k\in W^{3,2}_{-\kappa}(E_\rho )$ and our previous equation to get 
\[ 2(n^0l_k-l^0n_k)-(l^0+n^0)(l_k-n_k)=(n^0-l^0)(l_k+n_k)\in 
W^{3,2}_{-1-\kappa +6\epsilon}(E_\rho ).\]
Using lemma 1 and the asymptotic staticity of the system
\[ 2e\sum^3_{j=1}A^j(l^jn_k-n^jl_k)\in 
W^{2,2}_{-\kappa -1+2\epsilon}(E_\rho ).\] 
Putting this all together in the equation for $\partial_k\chi$ we have 
\[ \partial_k\zeta +\sqrt{2}\left[ (\varepsilon m+eA^0)\frac{1}{\sqrt{2}}
(l^0+n^0)-\varepsilon m\cos\zeta\,\right](l_k-n_k)\in 
W^{2,2}_{-1-\kappa +9\epsilon}(E_\rho ).\]
We need to refine our estimate for 
$(l^0-n^0)/\sqrt{2}$, noting 
\[ \left(\frac{l^0+n^0}{\sqrt{2}}\right)^2-1=\frac{1}{\sqrt{2}}
\sum^3_{j=1}(l^j+n^j)^2\in W^{3,2}_{-2\kappa}(E_\rho ),\]
and $\displaystyle{ [\frac{1}{\sqrt{2}}(l^0+n^0)+1]^{-1}\in 
W^{3,2}_{2\epsilon}(E_\rho )}$ we have 
\[ \frac{1}{\sqrt{2}}(l^0+n^0) -1\in W^{2,2}_{-2\kappa +2\epsilon}(E_\rho ).\]
Now use this estimate, together with lemma 1 (to separate the monopole term) 
and the fact that $(l_k-n_k) +2l^k\in W^{3,2}_{-\kappa}(E_\rho )$ (remember, 
$l_k =-l^k$), to get 
\[\partial_k\zeta -2\sqrt{2}\varepsilon m\left[1-\cos\zeta +
\frac{\varepsilon eq_0}{m|x|}+\frac{\varepsilon e}{m}a^0\right]l^k
\in W^{2,2}_{-\nu +9\epsilon}(E_\rho ),\]
where $\nu =\min [2\kappa , 1+\kappa]$. For convenience we will write this 
equation in a 3-vector notation, using $\nabla$ to denote the gradient and 
$l=(l^1,l^2,l^3)$; note that (because of asymptotic staticity) we have for the 
norm of $l$, $|l|^2=l.l=\frac{1}{2}$ plus a term in 
$W^{s,2}_{-\kappa +\epsilon}$ (see proof of lemma 2). We have,
\begin{equation}
\nabla\zeta -2\sqrt{2}\varepsilon m\left[1-\cos\zeta +
\frac{\varepsilon eq_0}{m|x|}+\frac{\varepsilon e}{m}a^0\right]l
\in W^{2,2}_{-\nu +9\epsilon}(E_\rho ).\label{eqn:zeta}\end{equation}

We will also require the asymptotically static version of equation 
(\ref{dR:inW}),
 the real part of equation (\ref{rchi:eq}). Firstly, from lemma 2 we have that 
\begin{eqnarray*}\sqrt{2}m^0& =&\sqrt{2}\sigma^0_{A\dot{A}}o^A\bar{\iota}\\
                    &=&-o_0\bar{\iota}^{\dot{1}}+o_1\bar{\iota}^{\dot{0}}\\
                    &\in & W^{3,2}_{-\kappa +2\epsilon}(E_\rho ).
\end{eqnarray*}
We have, after using, $(l_k-n_k) +2l^k\in W^{3,2}_{-\kappa}(E_\rho )$, 
\begin{equation}
\frac{1}{R}\nabla R +2\sqrt{2}\varepsilon m\sin\zeta\, l\in 
W^{2,2}_{-\min [1,\kappa]+4\epsilon}(E_\rho )\label{eqn:R}\end{equation}
Now, to the first of our two lemmas. In fact, this lemma actually gives us the 
exponential decay result which we use to improve lemma 1 so that we may obtain the much tighter estimate necessary for theorem \ref{thm:Emdecay}.
\begin{lemma} The Dirac field (and at least its first and second derivatives) 
of a 
stationary, isolated, asymptotically static (with $\kappa >1$, $s=3$ and 
$|E|=m$) Maxwell-Dirac field decays 
exponentially as $|x|\to\infty$, provided $q_0\neq 0$. In particular, 
\[ R<C\frac{e^{-2k\sqrt{|x|}}}{|x|},\,\,\,\mbox{for any $k$ such that}\,\,\,
0<k<2\sqrt{2}m\lambda ,\]
here $\lambda^2 = -\frac{\varepsilon eq_0}{m}$ is necessarily positive.
\label{lem:a}
\end{lemma}
{\em Proof} We first note from theorem 4 that $\sin\zeta\in 
W^{2,2}_{-\frac{1}{2}+4\epsilon}$ so, for $\rho$ large enough, 
$\zeta\in W^{2,2}_{-\frac{1}{2}+4\epsilon}(E_\rho )$ and $\nabla\zeta\in 
W^{1,2}_{-\frac{3}{2}+4\epsilon}(E_\rho )$.

Now, taking the divergence of equation (\ref{eqn:R}) and using 
(\ref{eqn:R}) again to 
remove the $|\nabla R|^2$ term,
\[ \frac{1}{R}\Delta R -4m^2\sin^2\zeta\in W^{1,2}_{-\frac{3}{2}+\epsilon}(
E_\rho )\,\,\,\mbox{for any}\,\,\, \epsilon >0.\]
From the fact that  $\nabla\zeta\in 
W^{1,2}_{-\frac{3}{2}+4\epsilon}(E_\rho )$ we have from (\ref{eqn:zeta}),
\[ 1-\cos\zeta\, +\frac{\varepsilon eq_0}{m|x|}+\frac{\varepsilon e}{m}a^0
=2\sin^2\frac{\zeta}{2}\, +
\frac{\varepsilon eq_0}{m|x|}+\frac{\varepsilon e}{m}a^0\in
W^{1,2}_{-\frac{3}{2}+4\epsilon}.\]
Our first observation is that $\varepsilon eq_0<0$, since from lemma 1 
$a^0\in W^{5,2}_{-\eta}$ with $\eta >1$. We write $\lambda^2=
-\frac{\varepsilon eq_0}{m}$, and take $\lambda >0$. 
We can also use the last inclusion to estimate the term $\sin^2\zeta$. We have,
\[\sin^2\zeta -\frac{2\lambda^2}{|x|}\in 
W^{1,2}_{-\min [\frac{3}{2},\eta ]+\epsilon}(E_\rho ).\]
The second order elliptic equation for $R$ can now be written as 
\begin{equation}
\Delta R -8m^2\left(\frac{\lambda^2}{|x|}+\alpha\right)R=0,\,\,\,\mbox{where}\,\,\, 
\alpha\in W^{1,2}_{-\min [\frac{3}{2},\eta ]+\epsilon}(E_\rho ).
\label{ell:R}\end{equation}
We will now use the maximum principle utilising a comparison function 
\[ v(x)=C\frac{e^{-2k\sqrt{|x|}}}{|x|},\,\,\, \mbox{for which}\,\,\, 
\Delta v -\left(\frac{k^2}{|x|}-\frac{k}{2|x|^{\frac{3}{2}}}\right)v=0.\]
Now,
\[\Delta [R-v(x)]-8m^2\left(\frac{\lambda^2}{|x|}+\alpha\right)[R-v(x)]=
\left[ (8m^2\lambda^2-k^2)\frac{1}{|x|}+\tilde{\alpha}\right]v(x),\]
where $\tilde{\alpha}\in W^{1,2}_{-\min [\frac{3}{2},\eta ]+\epsilon}(E_\rho )$.
Consequently, for $\rho$ large enough and for every $k$ such that 
$0<k<2\sqrt{2}m\lambda$ we have,
\[ \Delta [R-v(x)]-8m^2\left(\frac{\lambda^2}{|x|}+\alpha\right)[R-v(x)]>0.\]
Choosing $C$ such that $[R-v(x)]_{|x|=\rho}\leq 0$ we have by the maximum 
principle that $R-v(x)< 0$ on $E_\rho$. Completing the proof of our lemma.
\hfill{$\Box$}

\begin{lemma} For a stationary, isolated asymptotically static Maxwell-Dirac
system with $\kappa >1$, $s=3$ and $|E|=m$ we have the following estimates when $q_0\neq 0$.
\begin{eqnarray}
&&\zeta -\frac{\varepsilon_1\sqrt{2}\lambda}{\sqrt{|x|}}+
\frac{\varepsilon_1}{4m|x|}-\frac{\varepsilon_1(16\lambda^4m^2+9)}{96\sqrt{2}\lambda m^2 |x|^{\frac{3}{2}}}
\in W^{3,2}_{-2+\epsilon}(E_\rho );\nonumber\\
&&l=\frac{\varepsilon\varepsilon_1}{\sqrt{2}}\left(\hat{r}-u\right),\,\,\, 
\mbox{with}\,\,\, u\in W^{3,2}_{-\frac{1}{2}+\epsilon}(E_\rho ),\nonumber\\
&&\hat{r}.l-\frac{\varepsilon\varepsilon_1}{\sqrt{2}}\in W^{3,2}_{-1 +\epsilon}
\,\,\, \mbox{and}\,\,\, \hat{r}.u\in W^{3,2}_{-1 +\epsilon};\nonumber\\
&&\mbox{where}\,\,\, 
\hat{r}\,\,\,\mbox{is the radial unit vector and}\,\,\, 
(\varepsilon_1)^2=1.
\nonumber\label{expn:a}
\end{eqnarray}
\label{lem:b}\end{lemma}
{\em Proof} In all that follows we are assuming that $\rho$ is large 
enough that the necessary expansions -- eg $\sin\zeta -\zeta\in 
W^{2,2}_{-\frac{3}{2}+\epsilon}(E_\rho )$ when $\sin\zeta\in 
W^{2,2}_{-\frac{1}{2}+\epsilon}(E_\rho )$ -- can be made on $E_\rho$. 

We begin with the estimate, 
\[\sin^2\zeta -\frac{2\lambda^2}{|x|}\in 
W^{1,2}_{-\min [\frac{3}{2},\eta ]+\epsilon},\]
from the proof of lemma \ref{lem:a}. Write, 
\[ \zeta =\frac{\sqrt{2}\varepsilon_1\lambda}{\sqrt{|x|}}+\zeta_1\,\,\, 
\mbox{where}\,\,\, \varepsilon_1=\pm 1,\]
and substitute into equation (\ref{eqn:zeta})
\[ \nabla\zeta_1-\frac{\varepsilon_1\lambda}{\sqrt{2}|x|^{\frac{3}{2}}}\hat{r}
-2\sqrt{2}\varepsilon m\left[ \frac{2\sqrt{2}\varepsilon_1\lambda}{\sqrt{|x|}}
\zeta_1 +\zeta_{1}^{\phantom{1}2}\right]l\in W^{2,2}_{-2+\epsilon}(E_\rho ).\]
We have kept only terms ``less than order $1/|x|^2$" on the left of 
the equation. 
The $a^0$ from $A^0$ is of order $1/|x|^2$, since we can now improve the result 
of lemma 1 using lemma \ref{lem:a} -- from equation (\ref{max:is}) we have 
$\Delta A^0 =4\pi e\sqrt{2}(l^0+n^0)R$ so $A^0$ must be the sum of an harmonic 
polynomial (of negative degree) and a term which decays exponentially.

Starting with $\zeta_1\in W^{2,2}_{-\frac{1}{2}+\epsilon}$ (since $\zeta$ is) 
we have that $\nabla \zeta_1\in W^{1,2}_{-\frac{3}{2}+\epsilon}$, in the 
first instance. But then our equation (above) implies that $\zeta_1\in 
W^{1,2}_{-1+\epsilon}$ and that,
\[ \frac{1}{|x|}\hat{r}+4\sqrt{2}m\varepsilon\zeta_1\, l\in 
W^{-\frac{3}{2}+\epsilon}(E_\rho ).\]
Now write $\zeta_1 = -\frac{\varepsilon\varepsilon_2}{4m|x|}+\zeta_2$ 
($\varepsilon_2 =\pm 1$) and repeat
 the process to find that $\zeta_2\in W^{0,2}_{-\frac{3}{2}+\epsilon}$. As a 
consequence we have the following estimates for $\zeta$ and $l$, 
\begin{eqnarray}
&&\zeta -\frac{\sqrt{2}\lambda\varepsilon_1}{\sqrt{|x|}}+
\frac{\varepsilon\varepsilon_2}{4m|x|}
\in W^{0,2}_{-\frac{3}{2}+\epsilon}(E_\rho ),\nonumber\\
&&l=\frac{\varepsilon_2}{\sqrt{2}}\left(\hat{r}-u\right),\,\,\, 
\mbox{with}\,\,\, u\in W^{0,2}_{-\frac{1}{2}+\epsilon}(E_\rho ).\nonumber
\end{eqnarray}
Next we use the fact that, 
$l.l-\frac{1}{2}\in W^{3,2}_{-\kappa +\epsilon}$, as the system is 
asymptotically static. We have, 
\[ l.l-\frac{1}{2}=\frac{1}{2}(-2\hat{r}.u+|u|^2)\in W^{3,2}_{-\kappa +\epsilon}
.\]
It is a simple matter to show that for $u\in W^{3,2}_{2\epsilon}\cap 
W^{0,2}_{-\frac{1}{2}+\epsilon}$ we have $u\in W^{3,2}_{-\frac{1}{2}+\epsilon}$
:- 
Begin with a function $f$ in $W^{3,2}_{2\epsilon}\cap 
W^{0,2}_{-\frac{1}{2}+\epsilon}$ then integrate $\nabla .\left( \sigma^{2\delta 
-1}f\nabla f\right)$ over $E_\rho$ to show that $\nabla f\in 
W^{0,2}_{-1-\delta +\epsilon}$ for $0<\delta\leq\frac{1}{4}$, iterate this 
process to eventually find $\nabla f\in W^{0,2}_{-\frac{3}{2}+\epsilon}$ which 
together with $f\in W^{0,2}_{-\frac{1}{2}+\epsilon}$ shows that $f\in 
W^{1,2}_{-\frac{1}{2}+\epsilon}$. Now repeat the process with 
$\partial_k f\in W^{2,2}_{-1+\epsilon}\cap W^{0,2}_{-\frac{3}{2}+\epsilon}$, and
 so on to eventually get $f\in W^{3,2}_{-\frac{1}{2}+\epsilon}(E_\rho )$.
 With $u\in W^{3,2}_{-\frac{1}{2}+\epsilon}$, 
we can use the multiplication lemma to get 
$|u|^2=u.u\in W^{3,2}_{-1+\epsilon}$.Next we use the fact that,        
$l.l-\frac{1}{2}\in W^{3,2}_{-\kappa +\epsilon}$, as the system is      
asymptotically static. We have, 
\[ l.l-\frac{1}{2}=\frac{1}{2}(-2\hat{r}.u+|u|^2)\in W^{3,2}_{-\kappa +\epsilon}.\]
So that $\hat{r}.u\in W^{3,2}_{-1+\epsilon}$, as $\kappa >1$.

We also note that an argument similar to that used above 
shows that as $\zeta_2\in W^{3,2}_{\epsilon}\cap W^{2,2}_{-\frac{1}{2}+\epsilon}
\cap W^{0,2}_{-\frac{3}{2}+\epsilon}$ we must have
 $\zeta_2\in W^{3,2}_{-\frac{3}{2}+\epsilon}$. We can now substitute, 
\[\zeta_2=\frac{\alpha_0}{|x|^{\frac{3}{2}}}+\zeta_3,\,\,\, \mbox{where}\,\,\,
\alpha_0=\varepsilon_1\frac{(16\lambda^4m^2+9-96m^2d.\hat{r})}
{96\sqrt{2}\lambda m^2},\] 
here $d$ is a constant vector arising from the 
expansion of $\frac{\varepsilon e}{m}A^0=\frac{-\lambda^2}{|x|}+\frac{d.\hat{r}}
{|x|^2} +O(\frac{1}{|x|^3})$. We find that $\zeta_3\in W^{3,2}_{-2+\epsilon}$. 
We note from equation (\ref{eqn:zeta}) that $\hat{\theta}.\nabla\zeta$ and 
$\hat{\phi}.\nabla\zeta$ are both in $W^{3,2}_{-2+\epsilon}$ -- here 
$\hat{\theta}$ and $\hat{\phi}$ are the angular unit vectors orthogonal to 
$\hat{r}$. Consequently we must have $d=0$. 

 So, as $d$ gives rise to the 
electric dipole moment, we see that {\em the electric dipole moment must 
vanish.}

We still have to show that 
$\varepsilon_2 =\varepsilon\varepsilon_1$ to obtain the precise statements of 
the lemma. This is easily done by taking the estimates for $\sin\zeta$ and $l$ 
and substituting them into equation (\ref{eqn:R}) -- to obtain exponential 
decay (rather than growth!) we require $\varepsilon\varepsilon_1\varepsilon_2 
=1$.

\hfill{$\Box$}

\noindent{\em Proof of Theorem 5} Armed with lemma \ref{lem:b} the theorem is 
remarkably simple to prove. We start with equation (\ref{realc:2sp}), which can 
be written as 
\[ \frac{1}{R}l.\nabla R +\sqrt{2}\varepsilon m\sin\zeta\, +
\nabla .l =0.\]
Which gives,
\[ \frac{\hat{r}.\nabla R}{R}-\frac{u.\nabla R}{R}+2\varepsilon_1 m\sin\zeta +
\frac{2}{r}-\nabla .u =0.\]
Using equation (\ref{eqn:R}) we have (as $u.l\in W^{3,2}_{-1+\epsilon}$) 
\[ \frac{u.\nabla R}{R}\in W^{2,2}_{-\frac{3}{2}+\epsilon}.\]
Noting that $\nabla .u\in  W^{2,2}_{-\frac{3}{2}+\epsilon}$, we have, using 
lemma \ref{lem:b},
\[\frac{\hat{r}.\nabla R}{R}+2\sqrt{2}\frac{m\lambda}{\sqrt{|x|}}+\frac{3}{2|x|}
\in W^{2,2}_{-\frac{3}{2}+\epsilon}(E_\rho ).\]
Consequently, we have
\[ R=K\frac{e^{-4\sqrt{2}m\lambda\sqrt{|x|}+\beta}}{|x|^{-\frac{3}{2}}},\]
where $\beta\in W^{3,2}_{-\frac{1}{2}+\epsilon}$ and $K$ may depend on 
$\frac{x^i}{|x|}$. From equation (\ref{eqn:R}) we see that 
$\ln K\in W^{3,2}_{\epsilon}$ so that $K$ is a bounded function. The Sobolev 
inequality implies $|\beta |< C/|x|^{-\frac{1}{2}+\epsilon}$. The result now 
follows by bounding, $Ke^{\beta}$. 
\hfill{$\Box$}

\section{Discussion}

It is worth emphasising here that our results are based purely on rather weak
asymptotic regularity and decay assumptions. Nothing is assumed about the
behaviour of the fields in the interior region $B_\rho$.

Another important point to note is that all the results require the Maxwell 
equations only to obtain the decay conditions of lemma 1 and the improved decay 
required for theorem \ref{thm:Emdecay}. If this decay is 
given, {\em a priori}, then the results apply to the ``Dirac equation in an 
external field" as it is usually presented.

One question which needs to be addressed is the possible extension of the 
electric neutrality theorem of \cite{CH:pap3} to the asymptotically static 
case. But we leave this to a future paper.

\medskip

\section*{Appendix A: 2-Spinors and Spinor Dyads}

We collect here a number of facts relating to 2-spinor dyads and their 
associated null vectors. We give only a brief statement of the facts, for 
details the reader should consult the book of Penrose and Rindler, 
\cite{PR:book}.

\begin{itemize}
\item 2-spinor indices are raised and lowered with $\epsilon^{AB}$ and 
$\epsilon_{AB}$ (summation on repeated indices), $\displaystyle{\xi^A=
\epsilon^{AB}
\xi_B\,\,\,\mbox{and}\,\,\,
\xi_A=\epsilon_{BA}\xi^B\,\,\,\mbox{for any 2-spinor}\,\,\,\xi^A .}$
\item $\displaystyle{o_A\iota_B-o_B\iota_A =\epsilon_{AB}\,\,\,\mbox{and}\,\,\,
o^A\iota^B-o^B\iota^A =\epsilon^{AB}}$, where
\[ \left(\epsilon_{AB}\right) =\left(\epsilon^{AB}\right) = 
\left(\begin{array}{rr}0&1\\ -1&0\end{array}\right).\]
\item $o_A\iota^A=-o^A\iota_A=1$ and  $\bar{o}_{\dot{A}}\bar{\iota}^{\dot{A}}=
-\bar{o}^{\dot{A}}\bar{\iota}_{\dot{A}}=1$.
\item The van der Waerden symbols $\sigma^{\alpha}_{A\dot{A}}$ connect 
Minkowski vectors to 2-spinors and {\em vice versa}. The 
$\sqrt{2}(\sigma_k^{\phantom{k}A\dot{A}})$ (with $k=1,2,3$) 
are simply the Pauli matrices and 
$\sqrt{2}(\sigma_0^{\phantom{0}A\dot{A}})$ is 
the identity matrix. We have, 
$\sigma_{\alpha}^{A\dot{A}}\sigma_{\beta\,\,A\dot{A}}=\eta_{\alpha\beta}$ 
the Minkowski metric, and 
$\sigma_{\alpha\,\,A\dot{A}}\sigma^{\alpha}_{\phantom{\alpha}B\dot{B}}=
\epsilon_{AB}\epsilon_{\dot{A}\dot{B}}$.
\item Because of these relations the null vectors $l^{\alpha} = 
\sigma^{\alpha}_{A\dot{A}}o^A\bar{o}^{\dot{A}}$,
$n^{\alpha}= \sigma^{\alpha}_{A\dot{A}}\iota^A\bar{\iota}^{\dot{A}}$,
$m^{\alpha}= \sigma^{\alpha}_{A\dot{A}}o^A\bar{\iota}^{\dot{A}}$ and
$\bar{m}^{\alpha}= \sigma^{\alpha}_{A\dot{A}}\iota^A\bar{o}^{\dot{A}}$ form a
null tetrad; with $l_\alpha$ and $n_\alpha$ real, and $m_\alpha$
complex. We have, 
$l^\alpha l_\alpha =0$, $n^\alpha n_\alpha =0$, $m^\alpha m_\alpha =0$, 
$l^\alpha m_\alpha =0$, $l^\alpha\bar{m}_\alpha =0$, 
$n^\alpha m_\alpha =0$, $n^\alpha\bar{m}_\alpha =0$, 
$l^\alpha n_\alpha =1$ and $m^\alpha \bar{m}_\alpha =-1$.
\item For any vector $X_\alpha$ we have,
\[ X_\alpha = (n^\beta X_\beta )l_\alpha +(l^\beta X_\beta )n_\alpha -
(\bar{m}^\beta X_\beta )m_\alpha -(m^\beta X_\beta )\bar{m}_\alpha .\]
\item $\displaystyle{o^A\bar{o}^{\dot{A}}\partial_{A\dot{A}}f = l^\alpha
\partial_\alpha f}$, $\displaystyle{\iota^A\bar{\iota}^{\dot{A}}
\partial_{A\dot{A}}f = n^\alpha\partial_\alpha f}$, and so on.
\end{itemize}

\section*{Appendix B: Explicit Forms of the Dirac Equations}

In this appendix we collect together explicit forms of the Dirac and 
Klein-Gordon 
equations for stationary Maxwell-Dirac systems. 

In this section we use the notation $\displaystyle{ \partial_z =
\frac{\partial}{\partial z}\,\,\,\mbox{etc}\,\,\, , \partial =
\partial_x+i\partial_y}$, and 
for the electromagnetic potential $A^{\alpha}$, $A=A^1+iA^2$.

The Dirac bi-spinor is,
\[ \psi =e^{-iEt}\,\left(\begin{array}{c}
U_A\\ \bar{V}^{\dot{B}}\end{array}\right) .\]
The Dirac equations are,
\begin{eqnarray}
im(\bar{V}^{\dot{0}}-\frac{E}{m}U_0)-\bar{\partial}U_1-\partial_z U_0
-ie\left[ (A^0+A^3)U_0+\bar{A}U_1\right]=0&&\nonumber\hfill{(B.1)}\\
im(\bar{V}^{\dot{1}}-\frac{E}{m}U_1)-\partial U_0+\partial_z U_1
-ie\left[ AU_0+(A^0-A^3)U_1\right]=0&&\nonumber\\
im(U_1-\frac{E}{m}\bar{V}^{\dot{1}})+\partial\bar{V}^{\dot{0}}-
\partial_z\bar{V}^{\dot{1}}+ie\left[ -(A^0+A^3)\bar{V}^{\dot{1}}+
A\bar{V}^{\dot{0}}\right]=0&&\nonumber\\
im(U_0-\frac{E}{m}\bar{V}^{\dot{0}})+\bar{\partial}\bar{V}^{\dot{1}}+
\partial_z\bar{V}^{\dot{0}}+ie\left[ \bar{A}\bar{V}^{\dot{1}}-
(A^0-A^3)\bar{V}^{\dot{0}}\right]=0&&.\nonumber
\end{eqnarray}

The Klein-Gordon equations are easily derived via differentiation of 
equations (B.1), we give the results for $U$ only,
\begin{eqnarray}
&\Delta U_0+2ie\sum^3_{j=1}A^j\partial_jU_0+\left\{ (E^2-m^2)+
2eEA^0 +e^2A^{\alpha}A_{\alpha}\right.\phantom{++}&\hfill{(B.2)}\nonumber\\&+\left.ie\left[ 
\partial_z(A^0+A^3)+
\bar{\partial}A\right]\right\}U_0+ie\left[\partial_z\bar{A}+
\bar{\partial}(A^0-A^3)\right]U_1=0&\nonumber\\&&\nonumber\\
&\Delta U_1+2ie\sum^3_{j=1}A^j\partial_jU_1+\left\{ (E^2-m^2)+
2eEA^0 +e^2A^{\alpha}A_{\alpha}\right.\phantom{++}&\nonumber\\&+\left.ie\left[ 
-\partial_z(A^0-A^3)+
\partial\bar{A}\right]\right\}U_1+ie\left[-\partial_zA+
\partial (A^0+A^3)\right]U_0=0&\nonumber
\end{eqnarray}

Equation (\ref{emp:2sp}) gives the electromagnetic potential in terms of the 
$U$'s and $V$'s which may in turn be written terms of $R$, $\chi$ and the 
$o$'s and $\iota$'s. We give only the result for $A^0$,
\begin{eqnarray}
&A^0 =\frac{m}{2e}\left[ e^{-i\chi}(|o_0|^2+|o_1|^2+|\iota^0|^2+|\iota^1|^2)-
\frac{2E}{m}\right]\phantom{AAAAAA++AAAA}&(B.3)\hfill\nonumber\\ 
&+\frac{i}{2e}\left[ (\frac{\bar{\partial}R}{R}+
i\bar{\partial}\chi )\iota^0o_1 +\bar{\partial}(\iota^0o_1)+ (\frac{\partial R}{R}+i\partial 
\chi )\iota^1o_0 +\partial (\iota^1o_0)\right.\phantom{++}&\nonumber\\
&\left.+ (\frac{\partial_z R}{R}+i\partial_z 
\chi )\iota^0o_0 +\partial_z (\iota^0o_0)-(\frac{\partial_z R}{R}+i\partial_z 
\chi )\iota^1o_1 -\partial_z (\iota^1o_1)\right].&\nonumber
\end{eqnarray}

\end{document}